\documentclass[conference]{IEEEtran}
\IEEEoverridecommandlockouts
\usepackage{cite}
\usepackage{amsmath,amssymb,amsfonts}
\usepackage{algorithm}
\usepackage{algorithmic} 
\usepackage{graphicx}
\usepackage{textcomp}
\usepackage{xcolor}
\usepackage{bbm}
\usepackage{url}
\usepackage{balance}

\usepackage{varwidth}

\begin{document}

\title{
Representing 
Realistic Human Driver Behaviors using a Finite Size Gaussian Process Kernel Bank}

\author{Hossein Nourkhiz Mahjoub$^*$, Arash Raftari$^*$, Rodolfo Valiente$^*$, Yaser P. Fallah$^*$, Syed K. Mahmud$^{**}$ \\
$^*$Connected and Autonomous Vehicle Research Lab  (CAVREL) \\ Department of Electrical and Computer Engineering, University of Central Florida, Orlando, FL, USA\\
\{hnmahjoub, raftari, rvalienter90\}@knights.ucf.edu, yaser.fallah@ucf.edu \\
$^{**}$Hyundai America Technical Center, Inc. (HATCI) \\ Superior Township, MI, USA \\
smahmud@hatci.com
        
\thanks{This material is based on work supported in part by the National Science
Foundation under NSF CAREER Grant 1664968 and in part by the Hyundai America Technical Center, Inc. (HATCI).}}



\maketitle

\begin{abstract}
The performance of cooperative vehicular applications is tightly dependent on the reliability of the underneath Vehicle-to-Everything (V2X) communication technology. V2X standards, such as Dedicated Short-Range Communications (DSRC) and Cellular-V2X (C-V2X), which are passing their research phase before being mandated in the US, are supposed to serve as reliable circulatory systems for the time-critical information in vehicular networks; however, they are still heavily suffering from scalability issues in real traffic scenarios. The technology-agnostic notion of Model-Based Communications (MBC) has been proposed in our previous works as a promising paradigm to address the scalability issue and its performance, while acquiring different modeling strategies, has been vastly studied. In this work, the modeling capabilities of a powerful non-parametric Bayesian inference scheme, i.e., Gaussian Processes (GPs), is investigated within the MBC context with more details. Our observations reveal an important potential strength of GP-based MBC scheme, i.e., its capability of accurately modeling different driving behavioral patterns by utilizing only a limited size GP kernel bank. This interesting aspect of integrating GP inference with MBC framework, which has been verified in this work using realistic driving data sets, introduces this architecture as a strong and appealing candidate to address the scalability challenge. The results confirm that our proposed approach over-performs the state of the art research in terms of the required communication rate and GP kernel bank size.
 
\end{abstract}
\section{Introduction}
The last 20 years could be referred to as the age of Cooperative Vehicular Safety (CVS) Systems for the automotive industry. These life-saving systems were introduced after FCC allocated 75 MHz of spectrum at the 5.9 GHz frequency for the Intelligent Transportation Systems (ITS) in 1999. CVS systems have experienced a tremendous progress and improvement phase throughout these two decades. However, both of the underlying vehicular communication technologies which have been proposed and investigated so far, namely Dedicated-Short-Range-Communication (DSRC) and Cellular-Vehicle-to-Everything (C-V2X), are still challenging in terms of their scalability potential and can not completely satisfy the requirements of the time-critical vehicular safety applications in dense driving scenarios.\par
The scalability issue, which is due to the high percentage of packet drops in over-occupied and dense traffic situations, has been approached from different perspectives by researchers in vehicular academic and industrial community \cite{sae:j2945},\cite{jgozalvez:vtm},\cite{rtsim},\cite{controlsteering}. From a physical-layer perspective, different modulation and coding schemes (MCS) or specific power control methods have been introduced to address the problem \cite{sae:j2945}. In addition, different scheduling strategies have been tried in MAC layer to reduce the effect of packet collisions on safety-critical information exchange\cite{jgozalvez:vtm}. Some other works have also proposed different adaptive packet transmission rate algorithms which could be regarded as application layer remedies\cite{gbansal:limericacm}, \cite{ttielert:vnc}, \cite{yfallah:idmtvt}, \cite{clhuang:ieeenetwork}. However, all these different mechanisms are assuming the same content formation for each packet, based on SAE J2735 standard \cite{sae:j2735}, which has been commonly used by DSRC and C-V2X so far as the application layer packet dictionary set.

Recently, another paradigm, namely Model-Based Communications (MBC), has been proposed for the first time in \cite{yfallah:mbcsyscon} to address the scalability issue from a different point of view. This proposal, which  has been more elaborated in \cite{emoradipari:tiv2017}, \cite{hnmahjoub:vtc}, \cite{hnmahjoub:cavs}, \cite{hnmahjoub:IFAC19}, \cite{hnmahjoub:syscon19}, can be regarded as a \textit{PHY- and MAC-technology agnostic} scheme, designed from the application layer perspective. The core concept of MBC  proposes a new packet formation mechanism compared to the current Basic Safety Message (BSM) definition of the SAE J2735 standard.

Currently, different BSM fields defined by SAE J2735 are directly dedicated to GPS information read from an external GPS module in addition to raw information, such as velocity, longitudinal acceleration, steering wheel angle, etc., read from Controller Area Network (CAN). What MBC proposes is filling out the packet content in a more intelligent and efficient way by replacing the raw information with parameters of a mathematical forecasting stochastic model which encapsulates the combined stochastic driver/vehicle behavior and transmitting these parameters (and their updates as of needed) instead of disseminating raw information with a fixed or adaptive transmission rate.

Different modeling strategies have been explored in our previous works in order to find an appropriate scheme which could be integrated to our MBC architecture as its modeling sub-system. For instance, we have investigated  Hierarchical Dirichlet Process-Hidden Markov Model (HDP-HMM) \cite{hnmahjoub:vtc}, and Gaussian Processes (GPs)  \cite{hnmahjoub:cavs}, \cite{hnmahjoub:IFAC19}, \cite{hnmahjoub:syscon19}, which both fall under the Bayesian non-parametric inference umbrella. Our main rationale for pursuing this field of inference mechanisms, i.e., non-parametric Bayesian, to solve our problem is its fascinating and important capability to model different behaviors without imposing any prior restrictions and requirements on the behavior characteristics. This is necessary for our modeling framework, since it should be able to model any arbitrary and even unforeseen driver/vehicle behavior.

Among different non-parametric Bayesian methods, we found GP as a more appealing candidate and more aligned with our problem set up, based on the observations presented in our previous work \cite{hnmahjoub:syscon19}. In the aforementioned paper, the superiority of GP-based MBC framework over state-of-the-art baseline vehicular communication schemes has been demonstrated in terms of required message generation rate and also position tracking accuracy. Now, proceeding our previous research, we are trying to analyze the characteristics of the proposed GP-based MBC architecture in a more comprehensive way by looking at the specifications of the required GP kernel bank for modeling realistic driving behaviors. In order to achieve this goal, we selected the most representative trajectories from Safety Pilot Model Deployment (SPMD) data \cite{SPMD:Data}, which is a well-known realistic driving data set provided by US DOT, and tried to explore the properties of its required kernel bank, if it is modeled within GP-MBC framework. The details of this exploration is provided in the subsequent sections, but concisely, our findings show that the indirect vehicle position prediction with the required accuracy is feasible using just a finite-size set of GP kernels. Here what we mean by indirect position prediction is predicting the vehicle future positions using the Gaussian Process estimation of its future speed and heading time-series values. This observation is a very important and promising fact, since it enables us to achieve our final goal, i.e., deriving a Stochastic Hybrid System (SHS) modeling structure for driver behavior. In such a SHS framework, which this paper is an important milestone toward its realization, elements of the finite set GP kernel bank would serve as the representatives of different behavioral modes (discrete states of SHS).

The rest of this paper is organized as follows. In section \ref{Section:Problem_Statement} the overall framework is introduced in more details and its main building blocks, i.e. Gaussian Processes and Model-Based Communication are explained. Our experimental framework set up is illustrated in section \ref{Section:Experimental_Setup}. Section \ref{Section:Evaluation} is devoted to our results and observations, and finally section \ref{Section:Concluding_Remarks} concludes this work. 

\section{Problem Statement}

In our previous work, it has been demonstrated that within the model-based communication context, a hybrid modeling scheme which employs GP as its inference method has a notable superiority over the baseline modeling approach \cite{hnmahjoub:syscon19}. Here in this paper, we want to learn more about the characteristics of Gaussian Process kernels which are required to model the human driver behaviors within a GP-based MBC framework. However, before diving deeply into the core contributions of this work, it would be beneficial to overview the Model-Based Communication (MBC) concept and also the notion of Gaussian Process (GP) inference. 
This section is briefly describing our overall design and its core components, i.e., the \textit{error-driven MBC strategy} and \textit{GP inference}, which are essential for the rest of this work. Interested readers could refer to our previous works for further information \cite{yfallah:mbcsyscon}, \cite{emoradipari:tiv2017}, \cite{hnmahjoub:vtc}, \cite{hnmahjoub:cavs}, \cite{hnmahjoub:IFAC19}, \cite{hnmahjoub:syscon19}.

As mentioned in \cite{hnmahjoub:syscon19}, the overall framework of our work could be referred to as an error-driven model-based communication strategy for vehicular networks. The current standardized information encapsulation format for inter-vehicle communication in VANETs is wrapped in the so-called "SAE Dedicated Short Range Communications Message Set Dictionary" or SAE J2735 standard \cite{sae:j2735}. This dictionary defines different fields of BSM which is the information container fed by each vehicle into the network as a broadcast message and conveys its most updated dynamical information. Any other vehicle potentially will be able to track the BSM transmitter by receiving the consecutive instances of this information block over the network. SAE J2735, which is now commonly agreed by DSRC and C-V2X, only defines the BSM fields and other information dissemination details, such as transmission rate, transmission power, packet scheduling scheme, etc., are defined in other standards (such as SAE J2975/1 standard for DSRC).

\begin{figure*}[t]
\centering
 \includegraphics[width=.90\textwidth]{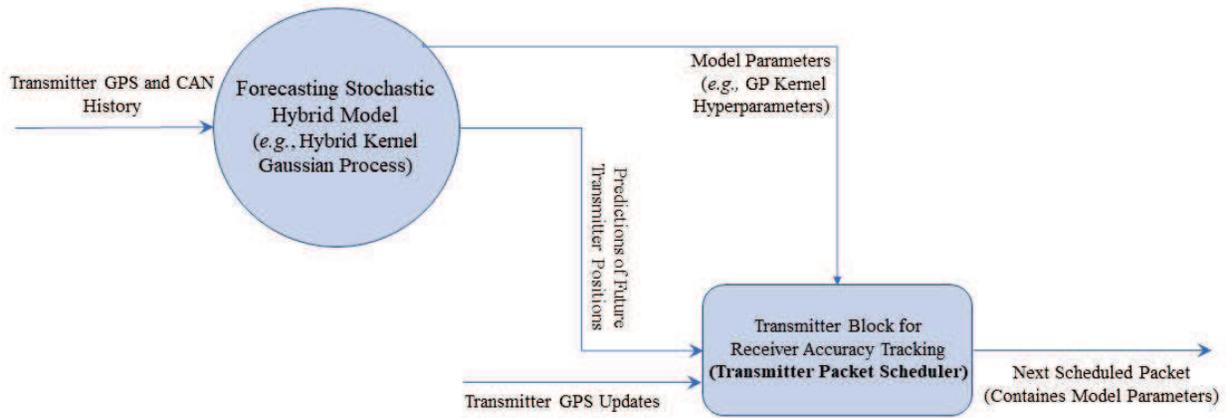}
    \caption{Error-Driven Hybrid Model-Based Communications Architecture}
    \label{Figure:ED-MBC_Architecture}
\end{figure*}

In \cite{hnmahjoub:syscon19} we combined the notion of model-based communication, which proposes an essentially different packet formation scheme compared to SAE J2735, with the error-driven packet scheduling approach, which is already part of the DSRC MAC layer standard, to build our proposed architecture for inter-vehicle information sharing in vehicular networks. This architecture is schematically depicted in Figure \ref{Figure:ED-MBC_Architecture}.

In this architecture, first the raw data is recorded from the GPS and CAN modules inside the host vehicle (packet transmitter). However, instead of using this raw information naively by pushing it directly into the BSM, our MBC design tries to find a mathematical forecasting model based on the observed history of this data and then fills out the BSM with the parameters of this model. This model is supposed to have notable longer prediction horizon compared to the base line constant speed model, which is currently the most dominant available prediction method in the commercial automotive industry. This longer prediction horizon, which will be referred to as "Model Persistency" from now on in this paper, potentially makes our architecture more robust against the packet drops and a more appealing and powerful solution to address the congestion challenge in dense driving scenarios.

\label{Section:Problem_Statement}

After finding an appropriate forecasting model, we take the error-driven approach to schedule the packet transmission instants. In this approach, which has been explained in details in \cite{hnmahjoub:syscon19}, there is a module inside the transmitter which tries to follow the receiver's tracking and estimation process on the fly to keep track of the receiver accuracy in estimating the transmitter's position. More specifically, at each GPS update this module simulates the receiver calculations for tracking the transmitter position assuming that the receiver has received the latest transmitter dynamics information (latest transmitted model update in MBC design) via the last BSM sent out by transmitter. This makes the transmitter capable of sitting on the receiver's seat and observing its own dynamics from the receiver's perspective. Transmitter then compares its estimation from the receiver tracking accuracy with a certain threshold, which normally is forced by the safety application requirements. Once the drift of estimated receiver accuracy from the actual transmitter dynamics exceeds this threshold, transmitter decides to schedule a new BSM (next model update in the MBC context) to update the receiver information.

According to the above explanations, it should be clear that the performance of this error-driven MBC design is highly dependent on the accuracy and strength of its underlying modeling scheme. Therefore, choosing a capable forecasting model is vital for our error-drive MBC architecture. As mentioned before, in our previous works several non-parametric Bayesian modeling methods, such as HDP-HMM and GP have been examined for this purpose. In general, non-parametric Bayesian methods allow us to infer the underlying behavioral patterns of observed time-series without imposing any presumptions or limitations on their characteristics. This phenomenal aspect relaxes the learning process from being bounded to specific function patterns. In other words, the complexity of a model which is derived in non-parametric Bayesian inference framework is automatically adapted to the observed data; and hence is able to both avoid creating over-complex models and capture the unforeseen patterns in the data on the fly.

Gaussian Process, which is one of the most powerful non-parametric Bayesian inference methods, has shown an enticing performance improvement in our recent studies in terms of required packet generation rate and also position tracking accuracy under network congestion \cite{hnmahjoub:syscon19}. Mathematical details of Gaussian Process inference has been explained in \cite{Rasmussen:GP}, but as a brief description, Gaussian Processes try to regress the observed time-series realizations by putting a prior distribution directly over the function space, instead of function parameters space, in a form that any finite subset of draws from this distribution represent a multivariate Gaussian random vector. This method by nature adapts the model complexity to the observed data and makes it capable of capturing different trends while they appear in training data.

Different patterns are essentially recognized through different GP kernel types. More precisely, each observation point is assumed to be drawn from a Gaussian random variable. However, these normal random variables are not independent and have temporal correlations with their predecessors and successors. This correlation models the temporal relation between observations and makes GP a powerful method to capture different patterns within time series.

The set of \textit{m} observed values are modeled as an \textit{m}-dimensional multivariate Gaussian random vector which could be defined using a mean vector of length \textit{m} and an \textit{m}-by-\textit{m} covariance matrix. This covariance matrix, or GP kernel, is the core asset by which GP recognizes the inherent behavior of time series through the observed time series history and forecasts its future. The core GP components could be mathematically formulated as follows. For more details one can refer to \cite{Rasmussen:GP}. 

\begin{equation}
    f(t) \sim gp (m(t), k(t,t'))
\end{equation}

\begin{equation}
    \{X_i\}_{i=1,2,...,m} = \{f(t_i)\}_{i=1,2,...,m} \sim \mathcal{N}(\overline{\mu},\,\Sigma)\ 
\end{equation}

\begin{equation}
    \overline{\mu} = m(t_i); \ \Sigma_{i,j} = k(t_i, t_j) \ \forall i,j \in \{1,2, ..., m\}
\end{equation}

Different time series patterns are captured via different kernel types.  In this work, based on our findings in \cite{hnmahjoub:syscon19}, a compound kernel of RBF and linear is utilized for our simulations.

In the following sections of this paper we take our study of GP-based MBC framework one step further towards our ultimate goal which is developing a comprehensive SHS design for driver behavior modeling and communicating it over the vehicular network. Here in this paper, we explore to see if and how can we develop the required set of GP kernels, or the \textit{GP Kernel Bank} as we call it throughout this paper, which be rich enough to cover different realistic human driver behaviors.

\section{Experimental Setup}
\label{Section:Experimental_Setup}
In this section,  we elaborate further on the data set and data pre-processing. Then, we define a baseline to compare the results. The section will be concluded with details of our implementation.

\subsection{Data Set}
 In this paper, we used the realistic driving data set provided by US DOT, Safety Pilot Model Deployment (SPMD) \cite{SPMD:Data}, which has been used in recent similar studies \cite{hnmahjoub:vtc}, \cite{hnmahjoub:syscon19}. This well-known data set has a wide variety of trips collected over different urban roads with different types of vehicles and drivers. This characteristic makes SPMD a comprehensive data set which represents a diverse range of driver/vehicle behaviours and maneuvers. SPMD data set is composed of information collected through two different settings of Data Acquisition Systems (DAS-1 and DAS-2) in Ann-Arbor, Michigan. These systems provide different in-vehicle information logged from CAN, such as longitudinal velocity and acceleration, yaw rate, steering angle, turning signal status and etc., along with the vehicle GPS information over the whole trip duration.

\subsection{Data Preprocessing}
In order to select a diverse set of trajectories, we ranked all the trips in the aforementioned data set based on our experimental criteria and then chose the 26 richest ones which had the most alignment with our analysis goals. All trips were ranked on the merits of trip duration, number of successful lane changes, number of aborted lane changes, number of times the vehicle used turning signal in a trip, number of times the vehicle stopped in a trip and standard deviation of vehicle's yaw rate, steering angle, GPS heading and longitudinal acceleration. These 26 richest trips, which include more than $170,000$ sample points (more than $17,000$ seconds of trajectories), were plotted and visually inspected to confirm the proper selection.

\subsection{Baseline}
As it was discussed in the previous section, non-parametric Bayesian schemes are capable of being adapted to different maneuvers and driving behaviours. It was shown in our previous works that these modeling schemes incorporate more complex model structures in comparison with the constant speed or constant acceleration models and could be considered as one of the state-of-the-art modeling schemes in the literature of vehicular society \cite{hnmahjoub:syscon19}. Therefore, for the evaluation purposes, a GP-based modeling scheme which directly predicts the position (X-ENU and Y-ENU), as what had been proposed in \cite{hnmahjoub:syscon19}, is considered as our first baseline model in this paper. This modeling approach is called \textit{GP-Direct} from now on and will be briefly explained in the following section. In \cite{hnmahjoub:syscon19}, the authors had also introduced a \textit{Hybrid} framework which combines GP-Direct scheme with a constant speed model. This framework is called  \textit{Hybrid GP-Direct} and considered as our other baseline method in this paper. 
\subsection{Implementation}
\begin{figure*}[t]
\centering
 \includegraphics[width=.90\textwidth]{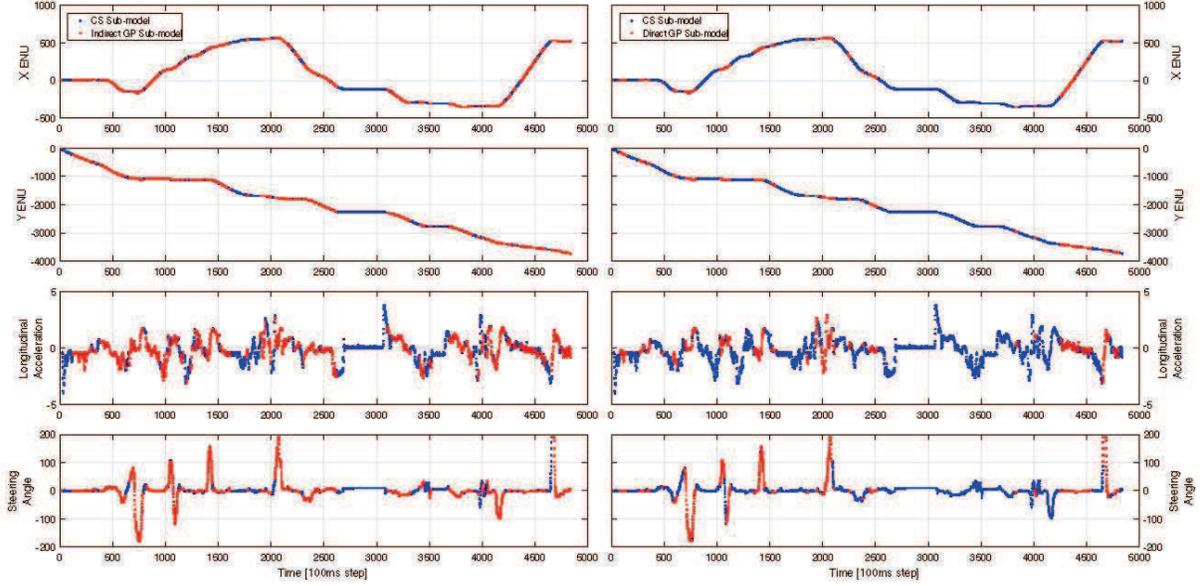}
\caption{Performance of the Hybrid GP modeling scheme in Direct vs. Indirect modes for ENU position prediction. Both schemes have the same threshold on the tracking error and are utilizing a two-state hybrid modeling structure, compromised of GP (with RBF + Linear kernel) and CV component. This figure (using different colors) shows the moments when a sub-model over-performs the other on the tracking accuracy while still remains below the threshold.}
\label{fig:hybrid_policy_comparison}
\end{figure*}

In our settings, GPS latitude, longitude and elevation have been converted into the East-North-Up (ENU) coordinate system. In the GP-Direct scheme, one of the baseline models in this work, X-ENU and Y-ENU are treated as two separate time-series, and are regressed by two GP models which are learned from their own histories. In the GP-Indirect modeling scheme instead of working directly on X-ENU and Y-ENU time series, heading and speed of the vehicle are treated as two independent time series which should be regressed using GPs. These GP models, which are learned from speed and heading histories, forecast their future value, and then predicted values of these two variables are used to predict position in ENU coordinate system. In both direct and indirect model generation schemes, a compound GP kernel type, composed of a linear and an RBF kernel, is selected based on our comprehensive observations. Results of these two different prediction schemes, i.e. Hybrid GP-Direct vs. Hybrid GP-Indirect, on a single trip is shown in Figure \ref{fig:hybrid_policy_comparison}. 

We conducted numerous experiments with different training window sizes, which is defined as the number of the latest equally spaced received samples (e.g., most recent history of ENU coordinates or heading and speed time series) utilized as the training data to generate each model. Then we chose the best training window and investigated four different position tracking error thresholds ($PTE_{th}$), i.e. 20 cm, 30 cm, 40 cm, and 50 cm. These values cover the range between minimum and maximum thresholds specified by SAE J2945/1 standard. In fact, position tracking error threshold determines the moments when the current model is not valid anymore and a new model should be selected from the available kernel bank or should be generated if an appropriate model for this data does not exist in the available kernel bank by this moment. The effect of position tracking error threshold on the persistency of models and the size of GP kernel bank are presented in the following section. The pseudo-code of our direct and indirect model generation schemes are illustrated in Algorithm \ref{mbc:algo} and Algorithm \ref{mbc:algo2}, respectively.

In \cite{hnmahjoub:syscon19}, the GP-Direct MBC algorithm was introduced; however, the existence of a finite size GP kernel bank which could be utilized to predict the position in different arbitrary trajectories with an acceptable level of error, according to a desirable \textit{PTE} threshold, has not been investigated so far. In this work, we investigate this hypothesis by introducing two model generation algorithms which designed to prevent adding redundant and repeated models to the kernel bank. The Algorithm \ref{mbc:algo} demonstrates the model generation procedure for the direct approach versus the indirect scheme of Algorithm \ref{mbc:algo2}.  As explained before, in the indirect scheme GP inference is employed to model speed and heading of the vehicle. After speed and heading are modeled, position of the vehicle, i.e., $X_{ENU}$ and $Y_{ENU}$, could be derived using the following simple equations:  

\begin{equation}
    \{S_i\}_{i=1,2,...,m} = \{f_{speed}(t_i)\}_{i=1,2,...,m} \sim \mathcal{N}(\overline{\mu_s},\,\Sigma_s)\ 
\end{equation}
\begin{equation}
    \overline{\mu_s} = m_s(t_i); \ \Sigma_{s_{i,j}} = k_s(t_i, t_j) \ \forall i,j \in \{1,2, ..., m\}
\end{equation}
\begin{equation}
    f_{speed}(t) \sim gp (m_s(t), k_s(t,t'))
\end{equation}

\begin{equation}
    \{H_i\}_{i=1,2,...,m} = \{f_{heading}(t_i)\}_{i=1,2,...,m} \sim \mathcal{N}(\overline{\mu_h},\,\Sigma_h)\ 
\end{equation}
\begin{equation}
    \overline{\mu_h} = m_h(t_i); \ \Sigma_{h_{i,j}} = k_h(t_i, t_j) \ \forall i,j \in \{1,2, ..., m\}
\end{equation}
\begin{equation}
    f_{heading}(t) \sim gp (m_h(t), k_h(t,t'))
\end{equation}
\begin{equation}
   X^{Predicted}(t_1) = X(t_0)  + \int_{t_0}^{t_1} f_{speed}(t)\, cos(f_{heading}(t))\, dt 
\end{equation}
\begin{equation}
   Y^{Predicted}(t_1) = Y(t_0)  + \int_{t_0}^{t_1} f_{speed}(t)\, sin(f_{heading}(t))\, dt 
\end{equation}

By obtaining actual position from GPS logs at any time instant, position tracking error (PTE) could be calculated as the 2D Euclidean distance between the actual and predicted vehicle positions:

\begin{equation}
   PTE= \sqrt{(Erorr_X)^2 +(Erorr_Y)^2 }
\end{equation}

Since PTE sampling is dependent on the availability of actual position updates, it can be done at most at the sampling rate of GPS updates, which is 10 Hz for SPMD dataset.

In both algorithms, vehicle's states (position for GP-Direct scheme, and position, speed and heading for GP-Indirect scheme) for all 
$N$ available trips ($T_j$ , $j=1, ..., N$)  are loaded one after another. At the beginning of each trip, $t_0$ (prediction start time) is initialized and the algorithm tries to predict the time series belong to this trip using the kernels which have been created so far. When a kernel is selected, its prediction accuracy is evaluated at each time-step ahead. As long as the kernel predicts the future positions with a $PTE$ less than the threshold ($PTE_{th}$), it remains as the selected kernel for our model and keeps this title until its prediction error exceeds the threshold ($PTE >PTE_{th}$). The size of the time interval ,in which the latest selected model remains in use, is called Model Persistency ($MP$). At this moment either another kernel from our kernel bank is selected for predicting the position or, if none of the available kernels could satisfy the $PTE_{th}$, a new one is created and added to the bank. Finally, we update $t_0$ to start the prediction with the updated kernel.


\begin{algorithm}[t]
    \begin{algorithmic}
        \caption{Direct framework model generation} \label{mbc:algo}
        \REQUIRE Trips for training $\mathcal{T}=\{T_{1}, ...,T_{m}\}$ ; $j = 0$ \\ 
        \WHILE {$j<m$} 
            \STATE $t_0\gets t_{start}$ ; $i = 0$\\
            \STATE \textbf{Read Data of trip ${T}_j$, load $\mathcal{S}_{ENU}$}\par
      
            \WHILE{$t_0<t_{end}$}
            {

                \WHILE{$(PTE<PTE_{th})$}
                {
                
                    \begin{varwidth}[t]{\linewidth}
                    
                        \STATE $i\gets i+1$; $t_{next}\gets t_0+i$ \par
                     \STATE \textbf{$MP = [t_0:t_{next}]$}
                      \STATE $PTE$ = GPeval(\textit{kernel}, $\mathcal{S}_{ENU}[MP]$)
          
                    \end{varwidth}
                }
                \ENDWHILE 
                
                   \FOR{$kernel \in K $}
                       
                       \STATE $PTE_k$ = GPeval(\textit{$kernel_k$}, $\mathcal{S}_{ENU}[MP]$)

                        \ENDFOR
                         \STATE $PTE_{min} =\textbf{min}(PTE_k)$
                         
                         \IF{$(PTE_{min}<PTE_{th})$}
                            \STATE\textbf{ load a previous kernel}
                     
                         \ELSE
                         
                           \STATE \textbf{$PH= [t_{next}-TW:t_{next}]$}
                            \STATE $kenel_{new}$=GPfit($\mathcal{S}_{ENU}[PH]$)\par
                         
                          \STATE \textbf{save a new kernel, update $K$} 
                         \ENDIF

                \par
                
                \begin{varwidth}[t]{\linewidth}
                   
                    \STATE $PTE_{min} \gets \infty$;  $t_0 \gets t_{next};$ $i \gets 0$\
                \end{varwidth}
            }
            \ENDWHILE
        \STATE $j\gets j+1$;\par
        \ENDWHILE 
    \end{algorithmic}
\end{algorithm}

\begin{algorithm}[t]
    \begin{algorithmic}
        \caption{Indirect framework model generation} \label{mbc:algo2}
        \REQUIRE Trips for training $\mathcal{T}=\{T_{1}, ...,T_{m}\}$ ; $j = 0$ \\ 
        \WHILE {$j<m$} 
            \STATE $t_0\gets t_{start}$ ; $i = 0$\\
            \STATE \textbf{Read Data of trip ${T}_j$ load $\mathcal{S}_{speed}$ ; $\mathcal{S}_{heading}$}\par

            \WHILE{$t_0<t_{end}$}
            {

                \WHILE{$(PTE<PTE_{th})$}
                {
                
                    \begin{varwidth}[t]{\linewidth}
                    
                        \STATE $i\gets i+1$; $t_{next}\gets t_0+i$ \par
                     \STATE \textbf{$MP = [t_0:t_{next}]$}
                  \STATE $[Speed, Heading]= $ 
                           \STATE GPeval$(\textit{kernel}$, $\mathcal{S}_{speed}[MP]$, $\mathcal{S}_{heading}[MP]$)
           \STATE $(X^{predicted}, Y^{predicted})$ =getXY($Speed,Heading$)
              \STATE $PTE$=getPTE($X^{predicted}, Y^{predicted}$)
                    \end{varwidth}
                }
                \ENDWHILE 
                
                   \FOR{$kernel \in K $}
                     
                    \STATE $[Speed, Heading]= $
                          \STATE GPeval$(\textit{kernel}$, $\mathcal{S}_{speed}[MP]$, $\mathcal{S}_{heading}[MP]$)
                           
                           \STATE $(X^{predicted}_k, Y^{predicted}_k)$ =getXY($Speed,Heading$)
                           
                           \STATE $PTE_k$=getPTE($X^{predicted}_k, Y^{predicted}_k$)

                        \ENDFOR
                         \STATE $PTE_{min} =\textbf{min}(PTE_k)$
                         
                         \IF{$(PTE_{min}<PTE_{th})$}
                        \STATE\textbf{ load a previous kernel}
                    
                         \ELSE
                        
                            \STATE \textbf{$PH= [t_{next}-TW:t_{next}]$}
                       \STATE ${kernel}_{new}$=GPfit($\mathcal{S}_{speed}[PH]$, $\mathcal{S}_{heading}[PH]$)\par
                         
                          \STATE \textbf{save a new kernel, update $K$} 
                         \ENDIF

                \par
                
                \begin{varwidth}[t]{\linewidth}
                   
                    \STATE $PTE_{min} \gets \infty$;  $t_0 \gets t_{next};$ $i \gets 0$\
                \end{varwidth}
            }
            \ENDWHILE
        \STATE $j\gets j+1$;\par
        \ENDWHILE 
    \end{algorithmic}
\end{algorithm}

\section{Evaluation}
\label{Section:Evaluation}
Two main metrics that any MBC system can be evaluated upon are the over-the-air packet length and required rate of message exchange. To address the first metric, we tried to investigate whether there is a limited size GP kernel bank that is capable of accurately modeling any behavioral driving/vehicle pattern. The existence of such a kernel bank empowers the transmitting entities to only send the ID of the kernel instead of kernel itself which consequently reduces the packet length. To address the other metric, we defined average model persistency as how long, on average, a model remains valid for prediction before being switched to another model. In this paper, we compare our proposed Indirect framework with the direct one, used as a baseline. The superiority of the Hybrid GP-Direct framework to the basic MBC, which only incorporates the constant speed model, has been demonstrated in \cite{hnmahjoub:syscon19}.

\begin{figure}[t]
\centering
\includegraphics[width=.48\textwidth]{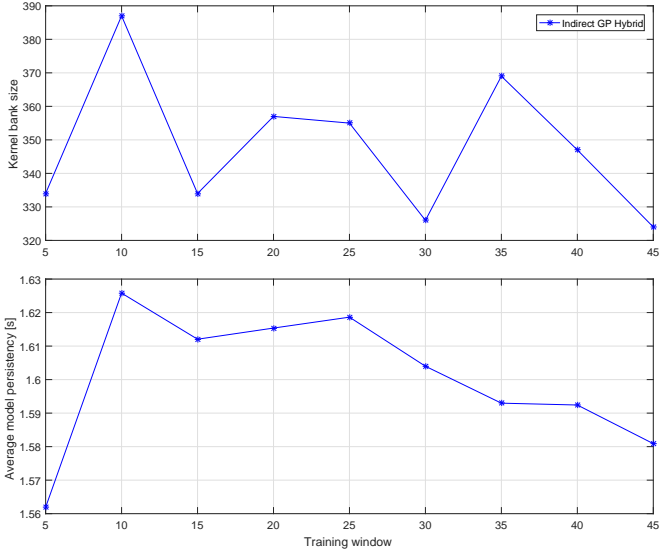}
\caption{Effect of training window size on the size of required kernel bank (Top) and average model persistency (Bottom) for Hybrid GP-Indirect method}
\label{fig:tws}
\end{figure}

First, the effect of training window size on the size of the kernel bank and persistency of models is investigated and the optimal size of training window is selected. For this purpose, we ran the model generation algorithm \ref{mbc:algo2} with different values for training window size.  Figure \ref{fig:tws} shows that the size of the training window has a negligible effect on average model persistency and size of the kernel bank. In fact, including more data of the past behaviour of the vehicle does not add more information for predicting the future short-term behavioural pattern. The best model persistency happened when only 10 most recent values of speed and heading were used for model generation; however, the size of required kernel bank is relatively bigger for training window of size 10. Based on these observations, the rest of experiments in this work are performed using training window size of 10 most recent samples. It also should be noted that in the presented results in this paper $PTE_{th}$ should be considered as 50 cm whenever its value is not mentioned explicitly.

\begin{figure}[t]
\centering
\includegraphics[width=.48\textwidth]{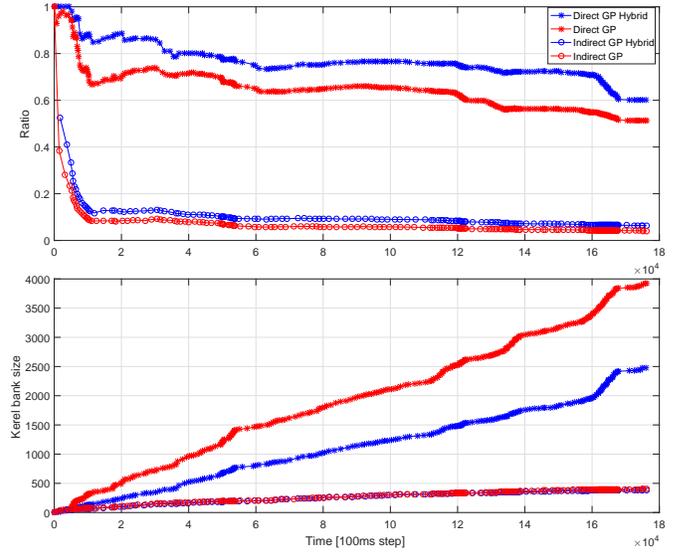}
\caption{ Ratio of model generation vs time (Top) and Kernel bank size vs time (Bottom)}
\label{fig:model-generation}
\end{figure}

Aiming at demonstration of our proposed framework capability to capture realistic human driver behaviors using a finite size GP kernel bank, we compare the baseline and our approach in terms of model generation. Figure \ref{fig:model-generation} illustrates the ratio of model generation (top) and the size of kernel bank (bottom) for both GP-Direct and GP-Indirect schemes at every moment. Model generation ratio at each moment is defined as the ratio of number of distinct kernels which have been generated so far (size of kernel bank at this moment) over the number of all moments when a model update is required due to inaccuracy of the currently in-use model (either this model update has been done by selecting a proper model from the kernel bank or by generating a new one). In Figure \ref{fig:model-generation} the Hybrid framework for both  schemes (GP-Direct and GP-Indirect) is shown in blue, while the solo GP-based (non-Hybrid) approach is shown in red. This figure clearly demonstrates that the model generation process is not converging for GP-Direct scheme and asymptotically tends to infinity. Therefore, we will eventually need infinite number of kernels to cover all possible driving behaviors in this case. On the contrary, for GP-Indirect framework, the model generation process shows a beautiful converging trend, which consequently indicates that the prediction of all behavioral patterns could be achieved by a limited-size kernel bank using GP-Indirect scheme.

\begin{figure}[t]
\centering
\includegraphics[width=.48\textwidth]{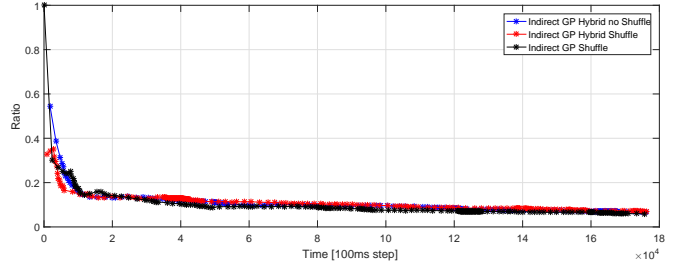}
\caption{ Effect of shuffling the training data on model generation ratio}
\label{fig:shuflle}
\end{figure}

In order to verify that this interesting capability of our proposed GP-Indirect scheme is not the result of richness and inclusiveness of the first trips in the training data, we repeated the same experiment on the shuffled version of our training data set. These shuffled versions are crafted by randomly rearranging the 26 trips in different orders. Figure \ref{fig:shuflle} presents the model generation ratio for two different shuffled versions of the data-set for both frameworks, i.e., GP-Indirect-Hybrid (in red) and GP-Indirect (in black), and the original order of training data for GP-Indirect-Hybrid (in blue). Similar figures have been observed for all other shuffled versions. Figure \ref{fig:shuflle} distinctly suggests that the model convergence is not dependent on the order of the trips in training process. 

\begin{figure}[t]
\centering
\includegraphics[width=.48\textwidth]{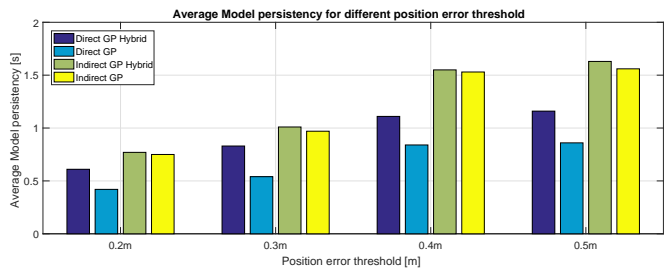}
\caption{Average model persistency for different choices of tracking error thresholds}
\label{fig:rate-comparison}
\end{figure}

\begin{figure}[t]
\centering
\includegraphics[width=.48\textwidth]{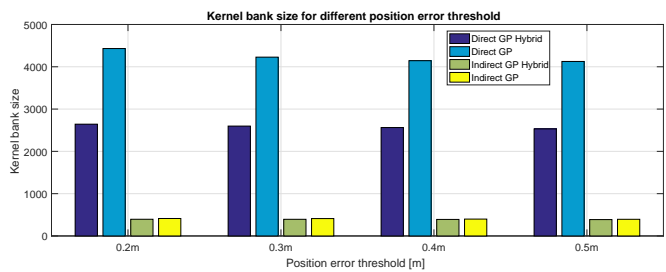}
\caption{Kernel bank size for different choices of tracking error thresholds}
\label{fig:rate-comparison2}
\end{figure}

Finally, to investigate the effect of position tracking error threshold constraint on model persistency and size of the kernel bank, we examined the model generation algorithm under four different threshold values recommended by SAE J2945/1. The average model persistency, shown in Figure \ref{fig:rate-comparison}, evidently suggests that compared to the baseline scheme, GP-Indirect scheme requires lower message exchange rate between transmitting entities in order to predict their positions with an acceptable error level. This phenomena can be attributed to the capability of GP-Indirect scheme to capture higher order vehicle dynamics resulting from hard brakes, lane change and turning maneuvers. Figure \ref{fig:rate-comparison2} shows that the size of the kernel bank in GP-Indirect scheme is not affected by changing the threshold constraint and is significantly smaller than the baseline. This interesting observation can be interpreted as the comprehensiveness of GP-Indirect scheme's kernel bank which makes it capable of covering almost all possible driving behavioural patterns. Table \ref{table:comp} and Table \ref{table:comp2} tabulate the data presented in Figure \ref{fig:rate-comparison} and Figure \ref{fig:rate-comparison2}, respectively.

\begin{table}[htb]
\caption{Average model persistency for different choices of tracking error thresholds}
\begin{center}
\begin{tabular}{c c c c c}
\hline
Model & \multicolumn{4}{c}{Position error threshold [m]}\\

     & 0.2 & 0.3 & 0.4 & 0.5 \\
 \hline
 \hline
  Direct GP Hybrid  & 0.61 & 0.83 & 1.10 & 1.16 \\

   Direct GP   & 0.42 & 0.54 & 0.84 & 0.86 \\
    Indirect GP Hybrid  & 0.77 & 1.01 & 1.55 & 1.63 \\
    Indirect GP & 0.75 & 0.97 & 1.53 & 1.56 \\
 \hline
 \hline

\end{tabular}
\end{center}
\label{table:comp}
\end{table}

\begin{table}[htb]
\caption{Kernel bank size for different choices of tracking error thresholds}
\begin{center}
\begin{tabular}{c c c c c}
\hline
Model & \multicolumn{4}{c}{Position error threshold [m]}\\

     & 0.2 & 0.3 & 0.4 & 0.5 \\
 \hline
 \hline
  Direct GP Hybrid  & 2643 & 2598 & 2563 & 2531 \\

   Direct GP   & 4431 & 4228 & 4144 & 4126 \\
    Indirect GP Hybrid  & 395 & 394 & 391 & 387 \\
    Indirect GP & 413 & 411 & 399 & 394 \\
 \hline
 \hline

\end{tabular}
\end{center}
\label{table:comp2}
\end{table}



\section{Concluding Remarks}
\label{Section:Concluding_Remarks}
The notable superiority of error-driven GP-based MBC designs compared to raw-information dissemination in terms of required communication rate and tracking precision motivates us to investigate the existence of a limited size GP kernel bank which is capable of predicting all possible driving behavioural patterns. Therefore, in this work we have proposed two different Hybrid GP-based modeling schemes, namely direct and indirect schemes, and compared their performance. A conspicuous improvement is observed using our indirect framework against the direct method in terms of the required size of the GP kernel bank, which is an indicator of exchanged message size, and also the average model persistency, which is an indicator of required transmission rate.
These observations motivate us to investigate the existence of natural and meaningful driving patterns and maneuvers corresponding to these models in our future research. Those patterns could be utilized for long horizon prediction purposes, i.e. maneuver and even driver intention prediction in the future.

\balance

\bibliography{0Syscon2018-main.bib}{}

\begin{thebibliography}{10}

\bibitem{sae:j2945}
Society of~Automotive Engineers~(SAE).
\newblock \text{SAE J945/1}.

\bibitem{jgozalvez:vtm}
R.~Molina-Masegosa and J.~Gozalvez.
\newblock Lte-v for sidelink 5g v2x vehicular communications: A new 5g
  technology for short-range vehicle-to-everything communications.
\newblock {\em IEEE Vehicular Technology Magazine}, 12(4):30--39, Dec 2017.

\bibitem{rtsim}
G.~{Shah}, R.~{Valiente}, N.~{Gupta}, S.~M.~O. {Gani}, B.~{Toghi}, Y.~P.
  {Fallah}, and S.~D. {Gupta}.
\newblock Real-time hardware-in-the-loop emulation framework for dsrc-based
  connected vehicle applications.
\newblock pages 1--6, 2019.

\bibitem{controlsteering}
R.~{Valiente}, M.~{Zaman}, S.~{Ozer}, and Y.~P. {Fallah}.
\newblock Controlling steering angle for cooperative self-driving vehicles
  utilizing cnn and lstm-based deep networks.
\newblock pages 2423--2428, 2019.

\bibitem{gbansal:limericacm}
John~B. Kenney, Gaurav Bansal, and Charles~E. Rohrs.
\newblock Limeric: A linear message rate control algorithm for vehicular dsrc
  systems.
\newblock pages 21--30, 2011.

\bibitem{ttielert:vnc}
T.~Tielert, D.~Jiang, Q.~Chen, L.~Delgrossi, and H.~Hartenstein.
\newblock Design methodology and evaluation of rate adaptation based congestion
  control for vehicle safety communications.
\newblock pages 116--123, Nov 2011.

\bibitem{yfallah:idmtvt}
Y.~P. Fallah, C.~Huang, R.~Sengupta, and H.~Krishnan.
\newblock Analysis of information dissemination in vehicular ad-hoc networks
  with application to cooperative vehicle safety systems.
\newblock {\em IEEE Transactions on Vehicular Technology}, 60(1):233--247, Jan
  2011.

\bibitem{clhuang:ieeenetwork}
C.~Huang, Y.~P. Fallah, R.~Sengupta, and H.~Krishnan.
\newblock Adaptive intervehicle communication control for cooperative safety
  systems.
\newblock {\em IEEE Network}, 24(1):6--13, Jan 2010.

\bibitem{sae:j2735}
Society of~Automotive Engineers~(SAE).
\newblock \text{SAE J2735}.

\bibitem{yfallah:mbcsyscon}
Y.~P. {Fallah}.
\newblock A model-based communication approach for distributed and connected
  vehicle safety systems.
\newblock In {\em 2016 Annual IEEE Systems Conference (SysCon)}, pages 1--6,
  April 2016.

\bibitem{emoradipari:tiv2017}
E.~Moradi-Pari, H.~N. Mahjoub, H.~Kazemi, Y.~P. Fallah, and
  A.~Tahmasbi-Sarvestani.
\newblock Utilizing model-based communication and control for cooperative
  automated vehicle applications.
\newblock {\em IEEE Transactions on Intelligent Vehicles}, 2(1):38--51, March
  2017.

\bibitem{hnmahjoub:vtc}
H.~N. {Mahjoub}, B.~{Toghi}, and Y.~P. {Fallah}.
\newblock A stochastic hybrid framework for driver behavior modeling based on
  hierarchical dirichlet process.
\newblock In {\em 2018 IEEE 88th Vehicular Technology Conference (VTC-Fall)},
  pages 1--5, Aug 2018.

\bibitem{hnmahjoub:cavs}
H.~N. {Mahjoub}, B.~{Toghi}, and Y.~P. {Fallah}.
\newblock A driver behavior modeling structure based on non-parametric bayesian
  stochastic hybrid architecture.
\newblock pages 1--5, Aug 2018.

\bibitem{hnmahjoub:IFAC19}
Hossein~Nourkhiz Mahjoub, Mohammadreza Davoodi, Yaser~P. Fallah, and Javad~M.
  Velni.
\newblock A stochastic hybrid structure for predicting disturbances in mixed
  automated and human-driven vehicular scenarios.
\newblock {\em IFAC-PapersOnLine}, 51(34):400 -- 402, 2019.
\newblock 2nd IFAC Conference on Cyber-Physical and Human Systems CPHS 2018.

\bibitem{hnmahjoub:syscon19}
H.~N. {Mahjoub}, B.~{Toghi}, S.~M.~O. {Gani}, and Y.~P. {Fallah}.
\newblock V2{X} system architecture utilizing hybrid gaussian process-based
  model structures.
\newblock In {\em 2019 IEEE International Systems Conference (SysCon)}, pages
  1--7, April 2019.

\bibitem{SPMD:Data}
Safety pilot model deployment dataset.
\newblock
  \url{https://www.its.dot.gov/factsheets/pdf/SafetyPilotModelDeployment.pdf}.
\newblock Accessed: 2019-11-29.

\bibitem{Rasmussen:GP}
Carl~Edward Rasmussen and Christopher K.~I. Williams.
\newblock {\em Gaussian Processes for Machine Learning}.
\newblock The MIT Press, 2006.

\end{thebibliography}
\bibliographystyle{unsrt}
\end{document}